\documentclass[12pt]{article}
\usepackage{graphicx}
\usepackage{amsmath}
\usepackage{amssymb}
\usepackage{amsthm}
\usepackage{latexsym}

\begin{document}


\title{\LARGE
{\bf The Scaling Limit Geometry of Near-Critical 2D Percolation}
}

\author{
{\bf Federico Camia}
\thanks{E-mail: f.camia@few.vu.nl}\\
{\small \sl Department of Mathematics, Vrije Universiteit Amsterdam}\\
\and
{\bf Luiz Renato G. Fontes}
\thanks{E-mail: lrenato@ime.usp.br}\\
{\small \sl Instituto de Matem\'atica e Estat\'istica, Universidade de S\~ao Paulo}
\and
{\bf Charles M.~Newman}
\thanks{E-mail: newman@courant.nyu.edu}\\
{\small \sl Courant Inst.~of Mathematical Sciences,
New York University}
}


\maketitle

\begin{abstract}
We analyze the geometry of scaling limits of near-critical 2D percolation,
i.e., for $p=p_c+\lambda\delta^{1/\nu}$, with $\nu=4/3$, as the lattice
spacing $\delta \to 0$.
Our proposed framework extends previous analyses for $p=p_c$, based
on $SLE_6$.
It combines the continuum nonsimple loop process describing the full
scaling limit at criticality with a Poissonian process for marking double
(touching) points of that (critical) loop process.
The double points are exactly the continuum limits of ``macroscopically pivotal"
lattice sites and the marked ones are those that actually change state as $\lambda$
varies.
This structure is rich enough to yield a one-parameter family of near-critical
loop processes and their associated connectivity probabilities as well as related
processes describing, e.g., the scaling limit of 2D minimal spanning trees.
\end{abstract}

\noindent {\bf Keywords:} scaling limits, percolation, near-critical,
minimal spanning tree, finite size scaling.

\section{Introduction} \label{intro}

A geometric analysis of the continuum scaling limit (where the lattice spacing
$\delta \to 0$) of critical two-dimensional percolation has been carried out
in recent years.
In the case of site percolation on the triangular lattice, this analysis, which
built on work of Cardy~\cite{cardy}, Aizenman~\cite{aizenman,aizenman1}
and Aizenman and Burchard~\cite{ab}, is now entirely rigorous.
First Schramm~\cite{schramm} focused on the critical percolation ``exploration
path" and identified the only plausible candidate for its scaling limit as chordal
$SLE_6$ (the Schramm-Loewner Evolution with parameter $\kappa=6$~\cite{schramm}).
Then Smirnov~\cite{smirnov} proved convergence of crossing probabilities for
the triangular lattice to the conformally invariant Cardy formulas~\cite{cardy}
and sketched an argument for the convergence of the exploration path to $SLE_6$
(see~\cite{cn2} for a detailed proof of this convergence) and finally Camia and
Newman constructed~\cite{cn1} a certain loop process and proved convergence to
it~\cite{cn2} of the ``full scaling limit" on the triangular lattice -- i.e.,
proved that the collection of the boundaries of all the (macroscopic) critical
clusters converges in distribution to that process of countably many continuum
nonsimple loops in the plane.

\begin{figure}[!ht]
\begin{center}
\includegraphics[width=8cm]{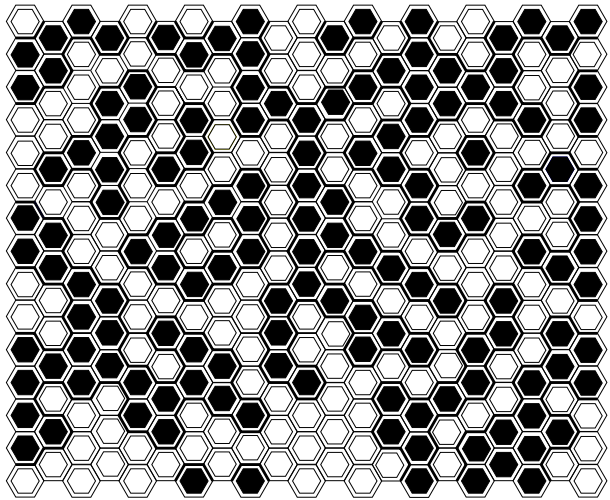}
\caption{Finite portion of a (site) percolation configuration
on the triangular lattice with each hexagon representing a
site assigned one of two colors.
In the critical percolation model, colors are assigned randomly
with equal probability.
The cluster boundaries are indicated by heavy lines; some small
loops appear, while other boundaries extend beyond the finite
region depicted.}
\label{percolation}
\end{center}
\end{figure}

\begin{figure}[!ht]
\begin{center}
\includegraphics[width=6cm]{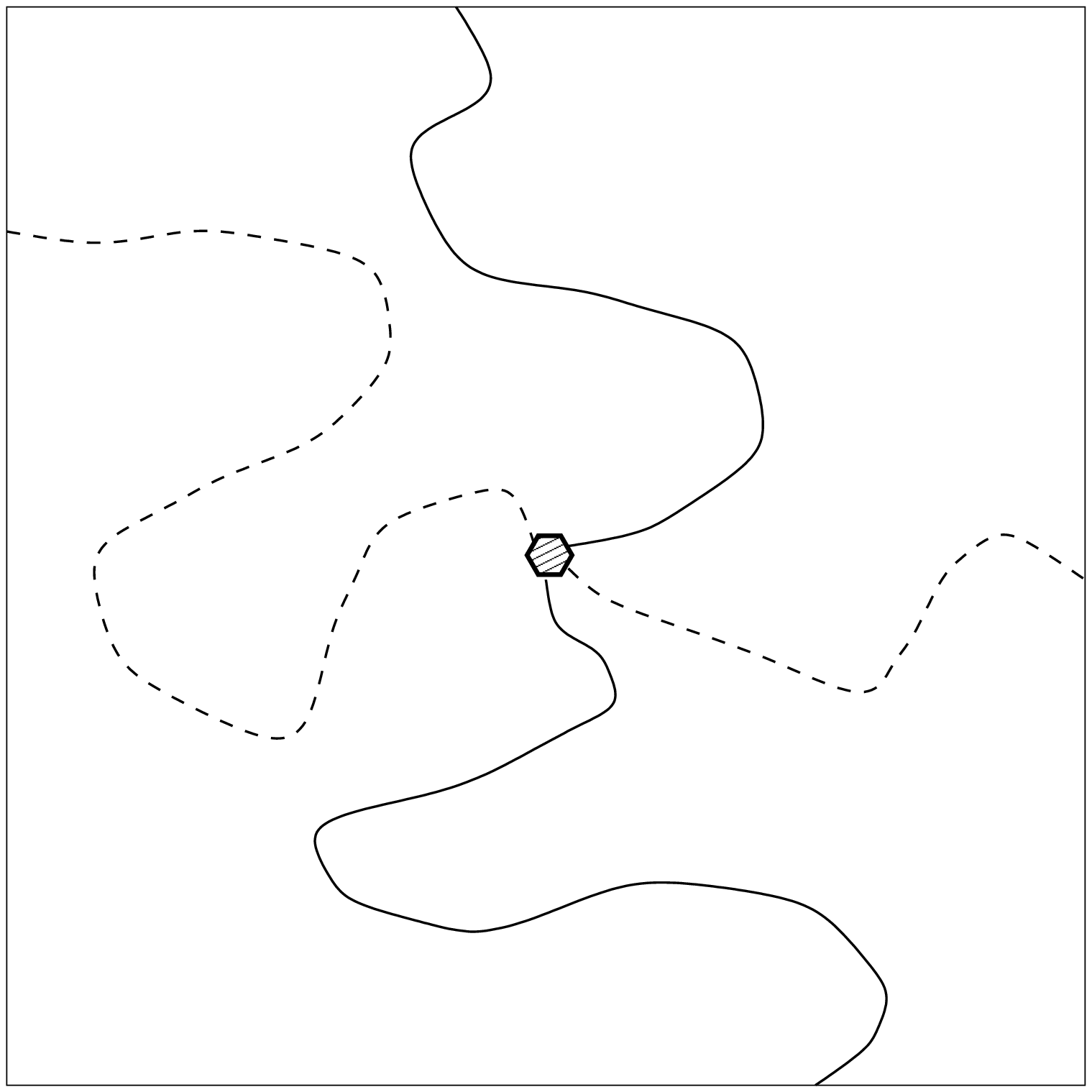}
\caption{Schematic diagram of a macroscopically pivotal hexagon at the center
of four macroscopic arms with alternating color. The full and dashed lines
represent paths of white and black hexagons respectively.}
\label{pivotal}
\end{center}
\end{figure}

Some properties of this critical loop process will be reviewed in Section~\ref{clp}.
Meanwhile, we point out one crucial feature that plays a key role throughout this
paper -- namely that although there is no self-crossing of a loop or crossing of
different loops, there is a considerable amount of self-touching of loops (i.e.,
the loops are non-simple) and touching between different loops.
These \emph{double points} of the loop process in the plane, as we will discuss in
Section~\ref{marking}, are exactly the continuum limits of ``macroscopically pivotal"
lattice locations; each such site (or bond, depending on the microscopic model in
question) is microscopic, but such that a change in its state (e.g., black to white
or closed to open) has a macroscopic effect on connectivity.
For site percolation on the triangular lattice (or equivalently random black/white
colorings of the hexagonal lattice -- see Figure~\ref{percolation}), a macroscopically
pivotal site is a hexagon at the center of four macroscopic arms with alternating
colors -- see Figure~\ref{pivotal}.

The critical value for triangular lattice site percolation (or square lattice
bond percolation) with probability $p$ of a site being white (or a bond being
open) is $p=1/2$.
The main purpose of this paper is to propose, and then analyze, a geometric
framework for scaling limits of \emph{near-critical} models where
$p=1/2+\lambda\delta^{\theta}$ as $\delta \to 0$ with $\lambda \in (-\infty,\infty)$
and $\theta$ chosen so that macroscopic connectivity functions in the scaling limit
have a nontrivial dependence on $\lambda$.
We note that scaling theory~\cite{stauffer} and the results of~\cite{bcks}
indicate that the correct choice is $\theta=1/\nu=3/4$, where $\nu$ is the
correlation length exponent.
Except for Section~\ref{mst}, we will focus on site percolation on the
triangular lattice, or equivalently, random colorings of the hexagonal lattice.

The analysis done in this paper is nonrigorous since our purpose here is not to
prove theorems but rather to propose a {\it conceptual\/} framework rich enough
to treat scaling limits of near-critical percolation and of related lattice objects
like the minimal spanning tree.
We hope however that our framework will be the foundation for an eventual detailed
rigorous analysis of near-critical and related two-dimensional scaling limits.

The framework we propose, based on a ``marking process" for the double points
of the critical ($\lambda=0$) full scaling limit of~\cite{cn1,cn2}, is presented
in Section~\ref{marking} below.
It provides a random marking of countably many double points, with each of these
labelled by a number in $(-\infty,\infty)$ representing the value of $\lambda$
at which that double point changes its state and hence correspondingly changes
macroscopic connections, loops, etc.
This yields a realization on a single probability space of all the scaling
limits as $\lambda$ varies in $(-\infty,\infty)$.
We point out that most double points are not marked since they do not change
their state for a finite value of $\lambda$ (in the limit $\delta \to 0$) --
it is only the marked ones that change.
We also note that an analogous, but simpler, marking procedure involving double
points of the ``Brownian web" has been used in~\cite{finr} to perturb around the
scaling limit of zero-temperature coarsening models in one plus one dimensional
space-time.

In Section~\ref{connectivity} we explain how the original full scaling limit
at the critical point (what we now call the $(\lambda=0)$-loop process) and
the marking of double points are together sufficient to yield the scaling limit,
simultaneously for all $\lambda$, of connectivity probabilities and cluster
boundary loops in the lattice model with $p=1/2+\lambda\delta^{\theta}$.
In Section~\ref{perc} we analyze the percolation transition for the continuum
model as $\lambda$ varies and give a description of what is seen for $\lambda \neq 0$
inside the critical scaling window, i.e., on a spatial scale (of the order
of one correlation length) where the system continues to look critical.
Finally, in Section~\ref{mst} we discuss another natural scaling limit that
should be constructible from the $0$-loop process combined with marked double
points -- namely, the continuum minimal spanning tree (MST).
This will be explored in more detail along with other scaling limits, such as
of the tree dual to the MST, the lattice filling curve that separates the two
trees, the $\lambda$-exploration path, invasion percolation, and dynamical
percolation~\cite{hps,ps,sst}, in another paper~\cite{cfn}.

At the discrete level, objects like the MST are most easily described for
bond percolation on the square lattice, so in the last section we will focus
on that microscopic model in our discussions.
On the other hand, the self-matching property of the triangular lattice,
such that the percolation process and its dual live on the same lattice,
makes that lattice particularly convenient to work with, so we will use
that discrete model in the rest of the paper.
It is also the case that all the \emph{rigorous} work about the scaling limit
of percolation is so far limited to site percolation on the triangular lattice.
Of course, because of universality, the choice of lattice should not be relevant
after the scaling limit is taken.

\section{The Critical Loop Process} \label{clp}

\subsection{General Features} \label{general}

At the percolation critical point, with probability one there is
no infinite cluster (rigorously proved only in two dimensions and
high dimension); therefore the percolation cluster boundaries form
loops (see Figure~\ref{percolation}).
We will refer to the continuum scaling limit (as the mesh $\delta$
of the lattice goes to zero) of the collection of all these loops
as the continuum nonsimple loop process; its existence and some
of its properties have been obtained in~\cite{cn1, cn2}.
We note that the cluster boundaries are naturally directed so
that, for example, following a boundary according to its direction,
white is to the left and black to the right.
This gives to the collection of all boundaries a nested structure
in which loops of opposite orientation alternate.
The limiting (as $\delta \to 0$) loops also have this property.

The continuum nonsimple loop process can be described as a
``conformally invariant gas" of loops, or more precisely,
a conformally invariant probability measure on countable
collections of continuous, nonsimple, noncrossing,
fractal loops in the plane.
The loops can and do touch themselves and each other many
times, but there is zero probability for the occurrence of
any triple points; i.e., no three or more loops can come
together at the same point, and a single loop cannot touch
the same point more than twice, nor can a loop touch a point
where another loop touches itself.

Any deterministic point $z$ in the plane (i.e., chosen
independently of the loop process) is surrounded by an
infinite family of nested loops with diameters going to
both zero and infinity.
Consequently, any two distinct deterministic points of the
plane are separated by loops winding around each of them.
However, any annulus about the deterministic point $z$ with
inner radius $r_1 > 0$ and outer radius $r_2 < \infty$
contains only a finite number $N(z, r_1, r_2)$ of loops
surrounding $z$.
Another important property of the loop process is that any two
loops are connected by a finite ``path'' of touching loops.

A continuum nonsimple loop process with the same distribution
as the full scaling limit of critical percolation can be constructed
directly by an inductive procedure in which each loop is obtained as
the concatenation of an $SLE_6$ path with (a portion of) another
$SLE_6$ path (see Subsection~\ref{single-loop}).
This procedure is carried out first in a finite region $D$ of the
plane, and then an infinite volume limit, $D \to {\mathbb R}^2$,
is taken (see~\cite{cn1,cn2}).


Two simple examples of the type of connectivity/crossing probabilities
that can be expressed in terms of the continuum nonsimple loop
process are given below.
The formulation of these connectivity/crossing probabilities
in terms of a conformally invariant loop process implies the conformal
invariance of such quantities (early discussions of scaling limits
of connectivity functions and of the consequences of conformal
invariance for such quantities are given in~\cite{aizenman,aizenman1}).
The examples will also highlight the natural nested structure of the
collection of percolation cluster boundaries in the scaling limit.

\begin{figure}[!ht]
\begin{center}
\includegraphics[width=6cm]{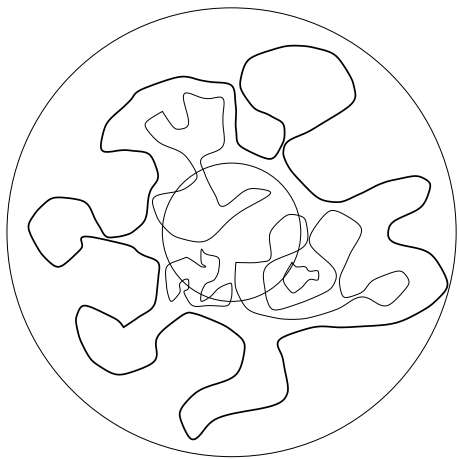}
\caption{An annulus whose inner disc is surrounded by a loop.
There is no monochromatic crossing between the inner and outer discs.
Other loops are shown in the figure, but they do not affect the
connectivity between the inner and outer discs.}
\label{annulus}
\end{center}
\end{figure}

Consider first an annulus centered at $z$ with inner radius
$r_1$ and outer radius $r_2$ (see Figure~\ref{annulus}).
The scaling limit $P(r_1,r_2)$ of the probability of a crossing
of the annulus (by crossing here we refer to a ``monochromatic"
crossing, i.e., a crossing by either of the two colors -- see
Figure~\ref{percolation}) can be expressed in terms of the loop
counting random variable $N(z,r_1,r_2)$ defined above: $P(r_1,r_2)$
is the probability that $N(z,r_1,r_2)$ equals zero.
More generally, $N(z,r_1,r_2)$ represents the scaling limit of the
minimal number of cluster boundaries traversed by paths connecting
the inner and outer circles of the annulus.

\begin{figure}[!ht]
\begin{center}
\includegraphics[width=8cm]{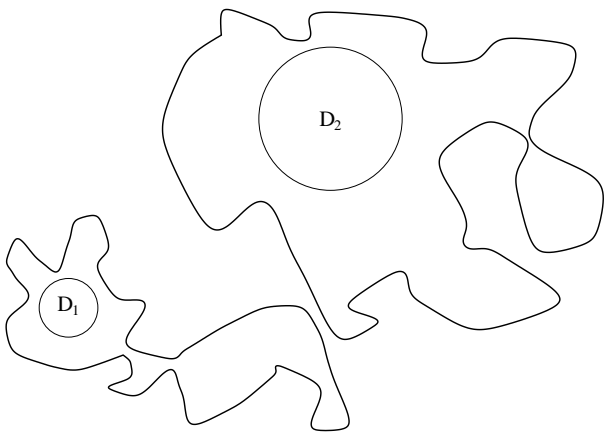}
\caption{Each one of the two disjoint discs in the figure is surrounded
by a loop that has the other disc in its exterior.
The minimal number of cluster boundaries that must be crossed to connect
the two discs is two.}
\label{discs}
\end{center}
\end{figure}

An example with more geometric structure involves two disjoint discs
$D_1$ and $D_2$ in the plane and the scaling limit $P(D_1,D_2)$ of
the probability that there is a crossing from $D_1$ to $D_2$ (see
Figure~\ref{discs}).
Here we let $N_1$ denote the number of distinct loops in the plane
that contain $D_1$ in their interior and $D_2$ in their exterior,
and define $N_2$ in the complementary way.
The scaling limit of the minimal number of cluster boundaries that
must be crossed to connect $D_1$ to $D_2$ is $N_1+N_2$, and $P(D_1,D_2)$
is the probability that $N_1=N_2=0$.
In the latter case, whether there is a white crossing between $D_1$
and $D_2$ or a black crossing or both will be discussed in
Section~\ref{connectivity} below.

One can also consider, as in~\cite{aizenman,aizenman1}, the probability
that a single monochromatic cluster in the exterior $E$ of the union of
$m$ disjoint discs (or other regions) connects all $m$ disc boundaries.
In the scaling limit, this can be expressed as the probability of the
event that there is a single continuous (nonsimple) curve in $E$
touching all $m$ disc boundaries that {\it does not cross any of the
loops of the continuum nonsimple loop process.}

We remark that the continuum nonsimple loop process described in this
section is presumably just one example of a family of ``conformal loop
ensembles" (see~\cite{werner3,ss,shw}) that are related to $SLE$ and to
the Gaussian free field, and are conjectured to describe the full scaling
limit of many statistical mechanics models besides percolation, such as
Ising, Potts and $O(N)$ models.

\subsection{Construction of a Single Loop} \label{single-loop}

\begin{figure}[!ht]
\begin{center}
\includegraphics[width=8cm]{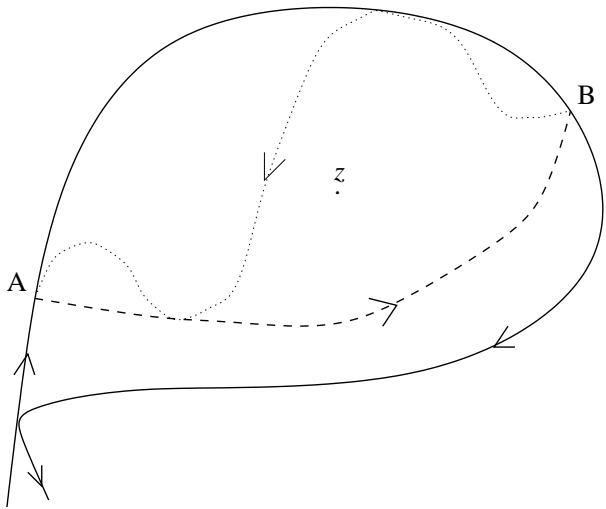}
\caption{Construction of a continuum loop around $z$ in three steps.
A domain $D$ is formed by the solid curve. The dashed curve is an
excursion $\cal E$ (from A to B) of an $SLE_6$ $\gamma$ in $D$ that
creates a subdomain $D'$ containing $z$. (Neither the rest of $\gamma$
nor its starting and ending points, $a$ and $b$, are indicated in the
figure.) The dotted curve $\gamma'$ is an $SLE_6$ in $D'$ from B to A.
A loop is formed by $\cal E$ followed by $\gamma'$.}
\label{fig-sec3}
\end{center}
\end{figure}

We will not give here the inductive construction of the full scaling limit
(see~\cite{cn1,cn2}), but in order to familiarize the reader with the loop
process, we explain in this subsection how to construct a single loop by
using two $SLE_6$ paths inside a domain $D$ whose boundary is assumed to
have a given orientation -- see Figure~\ref{fig-sec3}, where the orientation
is clockwise.
This is done in three steps, of which the first consists in choosing two
points $a$ and $b$ on the boundary $\partial D$ of $D$ and ``running'' a
chordal $SLE_6$, $\gamma(t) = \gamma_{D,a,b}(t), t \in [0,1]$, from $a$
to $b$ inside $D$.
We consider $\gamma[0,1]$ as an oriented path, with orientation from $a$
to $b$.
The set $D \setminus \gamma_{D,a,b}[0,1]$ is a countable union of its
connected components, which are each open and simply connected.
If $z$ is a deterministic point in $D$, then with probability one, $z$ is
not touched by $\gamma$~\cite{rs} and so it belongs to a unique domain in
$D \setminus \gamma_{D,a,b}[0,1]$.

The components of $D \setminus \gamma_{D,a,b}[0,1]$ can
be conveniently thought of in terms of how a point $z$ in
the interior of the component was first ``trapped'' at some
time $t_1$ by $\gamma[0,t_1]$, perhaps together with either
$\partial_{a,b} D$ or $\partial_{b,a} D$
(the portions of the boundary $\partial D$ from $a$ to $b$
counterclockwise or clockwise respectively):
(1) those components whose boundary contains a segment of
$\partial_{b,a} D$ between two successive visits at
$\gamma_0(z)=\gamma(t_0)$ and $\gamma_1(z)=\gamma(t_1)$ to
$\partial_{b,a} D$ (where here and below $t_0<t_1$), (2) the
analogous components with $\partial_{b,a} D$  replaced by the
other part of the boundary, $\partial_{a,b} D$, (3) those components
formed when $\gamma_0(z)=\gamma(t_0)=\gamma(t_1)=\gamma_1(z) \in D$
with $\gamma$ winding about $z$ in a counterclockwise direction
between $t_0$ and $t_1$, and finally (4) the analogous clockwise
components.

%

Now, let $D'$ be a domain of type 1 (if $\partial D$ were
counterclockwise, we would take a domain of type 2) and let
$A$ and $B$ be respectively the starting and ending point
of the excursion $\cal E$ that generated $D'$.
The second step to construct a loop is to run a chordal
$SLE_6$, $\gamma' = \gamma_{D',B,A}$, inside $D'$ from
$B$ to $A$; the third and final step consists in pasting
together $\cal E$ and $\gamma'$, keeping their orientations.

Running $\gamma'$ inside $D'$ from $B$ to $A$
partitions $D' \setminus \gamma'$ into new domains, all of
whose boundaries have a well defined orientation, so that
the construction of loops just presented can be iterated
inside each one of these domains (as well as inside each
of the domains of type~2, 3 and 4 generated by $\gamma_{D,a,b}$
in the first step).
For the complete inductive procedure generating all
the loops inside $D$, we refer the reader to~\cite{cn1, cn2}.

\section{Marking Pivotal Hexagons/Double Points in the
Lattice/Continuum} \label{marking}

The coupling (i.e., the realization on a single probability
space) of the $\lambda$-loop processes for 
all $\lambda \geq 0$ (with a symmetric picture applying for
$\lambda\leq0$) hinges on an ansatz about the evolution of
macroscopic loops at the lattice level as $\lambda$ increases
from $0$.
Note first that at the lattice level, there is a standard
way to couple all choices of $\lambda$ by means of i.i.d.
uniform $(0,1)$ random variables $U_h$ assigned to the hexagons,
labelled by $h$.
We then call a hexagon white for the $\lambda$-lattice model
(or more simply $\lambda$-white) if $U_h \leq 1/2 + \lambda \delta^\theta$,
so that a hexagon flips from black to white at the value of $\lambda$
where $1/2 + \lambda \delta^\theta$ crosses the value of $U_h$.
Note next that there are two ways in which a macroscopic loop will
change macroscopically in this setting: either by merging with another
macroscopic loop, or by splitting into two macroscopic loops.

The ansatz is that in both cases the change takes place by the
flipping to white of single black hexagons located either where
two distinct macroscopic boundary contours ``touch'' each other
(more precisely, come to distance $\delta$ from each other), or
where a single macroscopic boundary contour ``touches'' itself
(i.e., comes to a distance $\delta$ from itself).
In the first case the change from black to white will produce a
new macroscopic contour; in the second case two new contours are
formed (and an additional proviso of the ansatz is that both new
contours are macroscopic).
Each of the two cases has two subcases, as will be made more explicit
in Section~\ref{connectivity} in the context of the continuum model.


A hexagon at such a location is a black hexagon (at $\lambda=0$) which has
the macroscopic alternating four-arm property, requiring that there exist
four alternating (as one goes around the hexagon) white and black macroscopic
paths (of hexagons) touching the sides of the given hexagon.
Such hexagons, whether they be black or white, will be termed {\em important}.
The ansatz is that these are the only relevant hexagons, and other black
hexagons (at $\lambda=0$) do not play any (macroscopic) role when flipping to
white as $\lambda$ increases from $0$.
In particular, it should not be necessary to consider macroscopic changes
produced by the flipping of two or more hexagons,
each having just a microscopic effect on its own.
This is so because, as $\lambda$ increases, the probability of flipping
two or more hexagons is negligible compared to the probability of flipping
a single hexagon, and can be neglected, as long as the number of possible
pairs, triplets, etc. of hexagons whose coordinated flipping produces a
macroscopic effect is not significantly larger than the number of single
hexagons whose flipping has a macroscopic effect.

Suppose we denote by $N_k(\delta)$
the number of $k$-tuples (singlets,  pairs, triplets, etc.)
of hexagons in a fixed spatial region
whose simultaneous flipping produces a macroscopic change
(but the flipping of any $k-1$ of them does not do so).
Our ansatz relies on the hypothesis
that $N_k(\delta) = o(N_1(\delta)^k)$ for $k \geq 2$.
This is because the probability $p_k(\delta)$ of a significant
$k$-tuple flip will scale as $N_k(\delta)\, \delta^{\theta k}$
and so $p_1(\delta) = O(1)$ will imply that
$p_k(\delta) =o(1)$ for $k \geq 2$.

The flipping (black to white) of important hexagons will be governed
by the uniform random variables $U_h$ assigned to them.
An important hexagon is marked with a label $\lambda_0$ if the random
variable $U_h$ assigned to it equals $1/2+\lambda_0\delta^\theta$.
The exponent $\theta$ is chosen so that $N_1(\delta)$ scales
like $\delta^{-\theta}$; at the end of this section, we discuss why $\theta=3/4$.
In the limit $\delta \to 0$, the set of marked hexagons (or rather,
mark locations) along with their labels should converge in distribution
to a Poissonian point process in ${\mathbb R}^2 \times (0,\infty)$.
Moreover, and that is crucial, the set of marked important locations together
with the ensemble of ($\lambda=0$)-contours should converge jointly to the
above mentioned point process together with the continuum nonsimple loop
process, where the Poissonian nature of the point process is conditional
on the realization of the loop process.
At the discrete level the $\lambda$-marks are located on macroscopic
contours at touching points of two distinct such contours or at points
where a single such contour touches itself (with the additional proviso
mentioned earlier).
The same will hold in the continuum, namely the $\lambda$-marks are located
at touching points of two distinct loops or at points where a single loop
touches itself.

Now let us briefly describe the merging and splitting going on to form
$\lambda$-loops from continuum non-simple loops (corresponding to $\lambda = 0$)
together with the $\lambda$-marks.
A more complete discussion will be presented in Section~\ref{connectivity}.
Let us start with the merging of two counterclockwise continuum nonsimple
loops.
Suppose for a moment that these two loops are isolated from the rest of
the loop system, so that what we describe next makes proper sense.
Without recourse to this assumption, it makes sense only in a local way;
a global description is given in Section~\ref{connectivity}.
We look at the marked touching points of these two loops, and select the one
with the smallest $\lambda$-value, say $\lambda_0$, which will be strictly
positive by the scaling assumption (i.e., by the choice of $\theta$).
Then, as $\lambda$ increases from $0$ past $\lambda_0$, the two $0$-loops
merge into a single counterclockwise loop using the point with the
$\lambda_0$-value, which now becomes a point of the resulting loop where
it touches itself.

The splitting of a single counterclockwise loop is similar.
As in the previous paragraph, we make a simplifying (but presumably
technically incorrect) assumption to avoid global considerations -- 
in this case that among the marks appearing at the points where this 
loop touches itself there is a strictly positive smallest $\lambda$-value, 
say $\lambda_1$.
Then, as $\lambda$ increases from $0$ past $\lambda_1$, the loop splits
into two loops, one counterclockwise loop and a second one clockwise in
the interior of the counterclockwise loop, through the point with the
$\lambda_1$-value, which now becomes a point where the resulting loops
touch each other.

We end the section with an explanation of why $\theta=3/4$.
It relies on the above discussed correspondence of important points
to those having the four-arm property.
It is known that the probability that a unit macroscopic disk centered
at a given hexagon is such that the hexagon at the center has the four
arm property, with each arm touching the boundary of the disk, scales
like $\delta^{5/4}$.
This suggests that the number $N_1(\delta)$ of important hexagons
in a fixed macroscopic volume is of order
$\delta^{-2}\times \delta^{5/4}=\delta^{-3/4}$, and so in order to
obtain in the limit a process of important marked points, we should
scale the probability for a mark at each of the $O(\delta^{-3/4})$
important hexagons by $\delta^{3/4}$; thus $\theta=3/4$.
We note that this exponent is the same as the one of the scaling
window in~\cite{bcks}, where the scaling of the sizes of near-critical
large clusters is studied.
This is also the scale that makes the correlation length of order
$\delta^{-1}$, an observation that will be important later on (see Section~\ref{perc}).

\section{Continuum $\lambda_0$-Connectivity
and the $\lambda_0$-Loop Process} \label{connectivity}

In this section we show how the marking procedure discussed in the previous
section allows us to describe the scaling limit connectivity probabilities for
the one-parameter family of near-critical models with $p=1/2+\lambda\delta^\theta$.
As discussed in Section~\ref{clp}, in the critical case ($\lambda=0$),
two disjoint regions, $D_1$ and $D_2$, of the plane are connected (by a
monochromatic path) if there is no loop surrounding one region but not the other.
This is equivalent to saying that there is a continuous path from $D_1$ to $D_2$
that does not \emph{cross} any loops (although it can \emph{touch} loops).

\begin{figure}[!ht]
\begin{center}
\includegraphics[width=8cm]{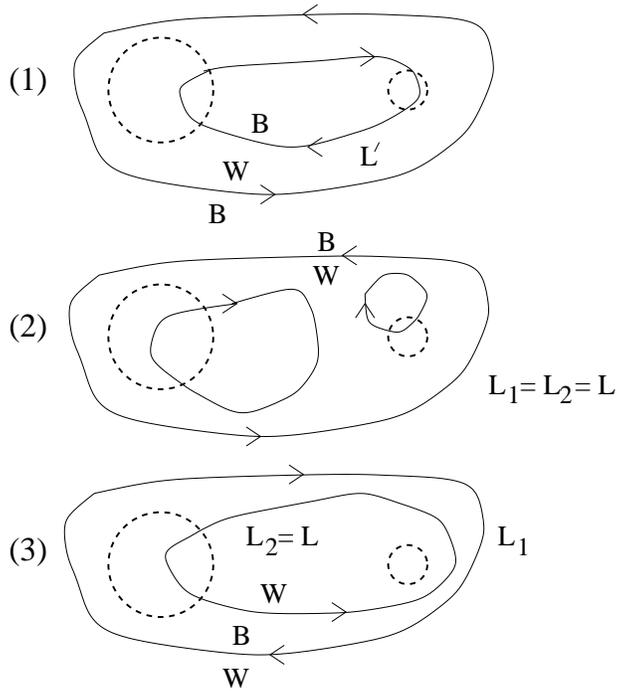}
\caption{Examples of monochromatic connections between the disc $D_1$ on the
left and the disc $D_2$ on the right. In (1), there are both white (W) and
black (B) connections; in (2) and (3), there is only a white connection.}
\label{connections}
\end{center}
\end{figure}

To determine whether there is a white connecting path or a black one or
both, let $L_i$ for $i=1,2$ denote the smallest loop surrounding $D_i$.
In the situation we are considering where there is a monochromatic
connecting path, there are three disjoint possibilities (see
Figure~\ref{connections}) -- either (1) there is a loop $L'$ touching
both $D_1$ and $D_2$, or (2) there is no such $L'$ and $L_1=L_2$, in
which case we define $L=L_1=L_2$, or (3) there is no such $L'$,
$L_1 \neq L_2$ and $L_2$ surrounds $L_1$ (resp., $L_1$ surrounds $L_2$),
in which case we define $L=L_1$ (resp., $L=L_2$).
Note that in case (3), $L_1$ touches $D_2$ (resp., $L_2$ touches $D_1$).
In case (1), there are both white and black connecting paths; in
cases (2) and (3), there is a white but no black connecting path
if $L$ is oriented counterclockwise, and otherwise there is a black
but no white connecting path.

The special case in which $D_1$ and $D_2$ are single points, $D_1=z_1$ and $D_2=z_2$,
is also included, but has to be treated with some care because if $z_1$ and $z_2$
are arbitrary deterministic points of the plane, the probability that they are
connected is zero.
Nevertheless, it may be instructive to think about this special case in
the lattice setting.
There, if the smallest boundary loop surrounding both $z_1$ and $z_2$ has
counterclockwise orientation and there is no boundary loop surrounding
either of the two points but not the other, then the hexagon containing
$z_1$ and that containing $z_2$ are in the same white cluster and thus
there is a path of white hexagons connecting them.

Now let us consider some $\lambda_0>0$ and look at the
$\lambda_0$-white connections.
Since this case corresponds to the scaling limit of models with
$p=1/2+\lambda_0\delta^\theta>1/2$, the white connectivity is ``enhanced"
and a $\lambda_0$-white path from $D_1$ to $D_2$ is \emph{allowed to cross}
critical ($\lambda=0$) loops, \emph{provided that it only crosses them
at marked sites with} $\lambda \leq \lambda_0$.
For every $\lambda_0 \geq 0$, this rule defines the set of all
$\lambda_0$-white paths and connections.
Using this definition, we can now consider the probability of connectivity
events of the type described in Section~\ref{clp} for the continuum
$\lambda_0$-percolation.
For example, given an annulus with inner radius $R_1$ and outer radius $R_2$,
we can ask for the probability of the event that there is a $\lambda_0$-white
crossing of the annulus.
This is the scaling limit, as $\delta \to 0$, of the corresponding
probability for the discrete percolation model with $p=1/2+\lambda_0\delta^\theta$.

The notion of $\lambda_0$-connectivity leads to the definition
of a $\lambda_0$-white cluster as a maximal set of points that
are connected by $\lambda_0$-white paths.
In the special case of $\lambda_0=0$, a continuum white cluster
can also be defined as the union of a counterclockwise $0$-loop
with its interior minus the interiors of all its daughter
(clockwise) domains.
This notion will appear again later, in Section~\ref{mst}.

The idea of $\lambda_0$-connectivity raises a natural question concerning
the percolation transition value for the continuum percolation model as
$\lambda_0$ varies, where percolation in this context means the existence
of a $\lambda_0$-white path to infinity.
From known properties of the critical ($\lambda_0=0$) model, namely
that every disc is surrounded by infinitely many loops, it follows
that there is no percolation for $\lambda_0=0$, so $\lambda_0=0$ is
a natural candidate for the transition value.
However, a priori, it could have happened that the probability of a
$\lambda_0$-white path to infinity is zero not only at $\lambda_0=0$,
but for all (or some) $\lambda_0 > 0$.
In the next section we will show that this is not the case and that the
probability of a $\lambda_0$-white path to infinity is strictly positive
(in fact, equal to one, if no starting region is specified) for every $\lambda_0>0$.

So far we have discussed $\lambda_0$-connectivity, but now we focus
on the related notion of $\lambda_0$-loops, i.e., the scaling limit
as $\delta \to 0$ of the collection of all boundary loops when
$p=1/2+\lambda_0\delta^\theta$.
By analogy with the critical ($\lambda_0=0$) case and the definition
of white paths there, the guiding idea is to define $\lambda_0$-loops
in such a way that $\lambda_0$-white paths \emph{do not cross}
any $\lambda_0$-loop.
In order to ensure this, one needs to merge together various $0$-loops
and split other $0$-loops, where the merging and splitting takes place
at marked double points and the decisions to merge or split depend on
the value associated to the mark and on $\lambda_0$.
The result of all this merging and splitting will be the collection
of $\lambda_0$-loops.

\begin{figure}[!ht]
\begin{center}
\includegraphics[width=8cm]{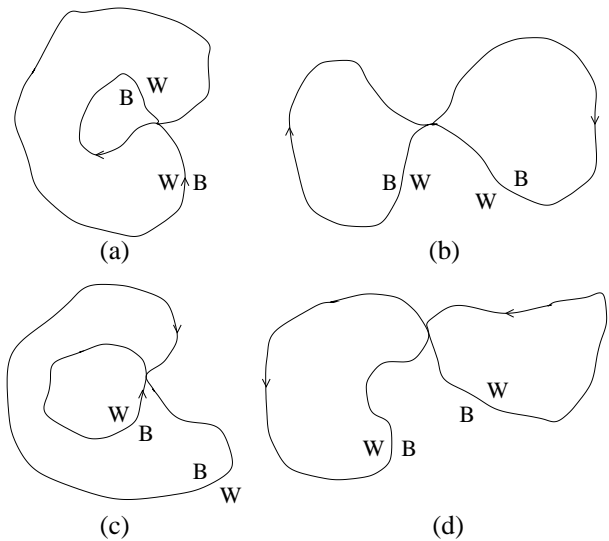}
\caption{Schematic diagram representing the different types of splitting and merging
caused by a black to white flip.
The arrows indicate the orientations of the loops, determined by having white (W)
on the left and black (B) on the right.}
\label{split-merg}
\end{center}
\end{figure}

For $\lambda_0>0$, the splitting of a $0$-loop (respectively,
merging of two $0$-loops), caused by a black to white flip, takes
place at a marked double point of the loop (respectively, where
the two loops touch) with $\lambda \in [0, \lambda_0]$.
There are two types of splitting and two types of merging
(see Figure~\ref{split-merg}):
\begin{itemize}
\item[(a)] the splitting of a counterclockwise loop into an outer counterclockwise
and an interior clockwise loop,
\item[(b)] the splitting of a clockwise loop into two adjacent clockwise loops,
\item[(c)] the merging of a counterclockwise loop with the smallest clockwise loop
that contains it into a clockwise loop, and
\item[(d)] the merging of two adjacent counterclockwise loops into a counterclockwise
loop.
\end{itemize}
The case $\lambda_0<0$ is of course exactly symmetric to the one described
here with splittings and mergings caused by white to black flips.

Before concluding this section, we point out that things are more complex
than they may first appear, based on the previous discussion.
This is because the critical loop process is scale invariant and each
configuration contains infinitely many loops at all scales, which implies
that in implementing the merging/splitting, one needs in principle both a
small scale $\varepsilon$-cutoff and a large scale $L$-cutoff.
This means that the merging/splitting is first done only for loops touching
a square centered at the origin of side length $L$ and takes place only if
both loops involved in the merging or both loops resulting from the splitting
have diameter larger than $\varepsilon$.
For every $0<\varepsilon<L<\infty$, this ensures that the number of merging
and splitting operations is finite.
At a later stage one takes both $L \to \infty$ and $\varepsilon \to 0$.

\section{Critical Scaling Window and $\lambda_0$-Percolation} \label{perc}

In this section we give a description of the system with
$p=1/2+\lambda_0\delta^\theta$ and $\lambda_0>0$, within the
spatial scaling window where it looks critical.
As a result of this analysis, we will answer the question regarding
continuum $\lambda_0$-percolation, as anticipated in the previous section.

First, we note that the order of magnitude of the linear dimension
of the critical scaling window is given by the correlation length
(measured in lattice units)
\begin{equation}
\xi(p) \approx (p-1/2)^{-\nu} = \lambda_0^{-\nu} \delta^{-\theta\nu},
\end{equation}
where $\nu=4/3$ is the correlation length exponent.
An important consequence of the choice of $\theta=3/4$ is that
$\xi(p)$ is of order $1/\delta$ as $\delta \to 0$.

Considering the percolation probability  on the lattice that the origin
belongs to an infinite white cluster, $\theta(p)$, we can write
\begin{equation} \label{theta}
\theta(p) \approx (p-1/2)^\beta \approx \delta^{\theta\beta} = \delta^{5/48},
\end{equation}
where $\beta=5/36$.
We thus see the one-arm exponent $1/\rho=5/48$ appear, with $\delta^{1/\rho}$
giving the order of magnitude  of the ${\mathbb P}_{p=1/2}$-probability
(i.e., calculated with $p=1/2$) that in the critical case there is a white
path starting at the origin and extending all the way to the boundary of
the disc of radius $1/\delta$ (measured in lattice units).
Since within distances of the order of one correlation length the
$\lambda_0$-system looks critical
and so we may estimate probabilities using ${\mathbb P}_{p=1/2}$, the
interpretation of~(\ref{theta}) is that in order to percolate it is
``sufficient" to reach the boundary of the disc of radius $1/\delta$
(measured in lattice units).

We can also interpret~(\ref{theta}) as meaning that the smallest
(in the sense of surrounding the smallest region) white circuit
$\cal C$ (i.e., a self-avoiding circuit formed out of white hexagons)
around the origin that belongs to the infinite cluster is at distance
of order $1/\delta$ (measured in lattice units)
from the origin.
Therefore, in macroscopic units (which are of order $1/\delta$ lattice
units), $\cal C$ stays at distance $O(1)$ as $\delta \to 0$.
In other words, the choice of the exponent $\theta$ is such that
$\cal C$ neither approaches the origin nor recedes to infinity in
the scaling limit.
This implies that, after the scaling limit has been taken, there is
a largest $\lambda_0$-loop around the origin, whose outer envelope is
at distance $O(1)$ from the origin.
Therefore, as anticipated in the previous section, for any $\lambda_0>0$,
the probability of the existence of a $\lambda_0$-white path to infinity
is strictly positive (in fact equal to one), which means that $\lambda_0=0$
is indeed the transition value for the continuum $\lambda_0$-percolation model.

For any $\lambda_0>0$, it is only within a distance $O(1)$ of the origin
that the continuum $\lambda_0$-percolation model looks critical.
In particular, the largest $\lambda_0$-loop that surrounds the origin is the
outer boundary of a $\lambda_0$-black cluster and has (the continuum
limit of) $\cal C$ as its outer envelope; the infinite $\lambda_0$-white
cluster has this largest $\lambda_0$-loop about the origin as one of its
inner boundaries.
From $\cal C$, various ``dangling ends'' of the infinite cluster
extend inside the critical scaling window coming closer to the origin.
If we focus only inside a fixed window (taken with free boundary
conditions), some of those dangling ends get disconnected from each
other and become separate clusters.
These are typically among the largest clusters inside the critical
scaling window, whose sizes are of order
$(1/\delta)^{2-1/\rho}=(1/\delta)^{2-\theta\beta}=\delta^{-91/48}$,
as has been rigorously proved in~\cite{bcks}.

\section{The Continuum Minimal Spanning Tree} \label{mst}

In this section, we propose a construction of the scaling limit of the
discrete minimal spanning tree (MST), using our framework of continuum
nonsimple loops and marked double points.
The discrete MST is most easily defined on the square lattice, so in
this section we focus on bond percolation on ${\mathbb Z}^2$.
For each nearest neighbor bond (or edge) $b$, let $U_b$ be a uniform $(0,1)$
random variable with the $U_b$'s independent.
This provides a standard coupling (i.e., realization on a single probability
space) of bond percolation models for all values $p$ of the probability that
a bond $b$ is open by saying that $b$ is $p$-open if $U_b \leq p$.
One then defines the minimal spanning tree in, say, an $L \times L$ square
$\Lambda_L$ centered at the origin as the spanning tree in $\Lambda_L$
with the minimum value of $\sum_b U_b$ summed over $b$'s in the tree.
It is known, based on the relation to invasion percolation, that there is
with probability one a single limiting tree as $L \to \infty$~\cite{ccn,ns}.
We will denote this tree on the $\delta$-lattice, $\delta{\mathbb Z}^2$,
by $T_{\delta}$.

The purpose of this section is to describe the putative scaling limit
(in distribution) of $T_{\delta}$ as $\delta \to 0$, in terms of our
critical $0$-loop process and $\lambda$-marked double points.
Our description uses a minimax construction in the continuum which is
a natural analogue of a well-known one on the lattice (see, e.g.,
\cite{alexander} and references therein).
We ignore differences between bond percolation on the square lattice and
site percolation on the triangular lattice in the belief that they have
no effect on the continuum scaling limit; in particular we will use
``white" and ``open" interchangeably.

%

In the scaling limit, we may consider the continuum MST as the limiting
set of paths  within the tree (see~\cite{abnw} for a general discussion
of continuum scaling limits of trees).
In order to define this tree, it is enough to describe the (unique, with
probability one) tree path between any two given deterministic points in
${\mathbb R}^2$.
However, it will be more convenient to describe the continuum tree path
between pairs of (non-deterministic) points, $z_1,z_2$, such that each
is contained in a continuum white cluster.
Since such points are dense in ${\mathbb R}^2$, one should obtain from
these the paths between all pairs of points, including deterministic ones.

For this purpose, we will use the idea of $\lambda$-connectivity
introduced in Section~\ref{connectivity}.
Any two points $z_1,z_2$ contained in continuum white (open) clusters
(of the critical model) are $\lambda$-connected for some large enough
value $\lambda<\infty$.
To find the tree path between $z_1$ and $z_2$, we start decreasing
$\lambda$ from $+\infty$ until it reaches a value $\lambda_1$ below
which $z_1$ and $z_2$ are not $\lambda$-connected.
$\lambda_1$ is the smallest $\lambda$ for which $z_1$ and $z_2$ are
$\lambda$-connected and furthermore $\lambda_1$ is the value of a
unique marked point $\zeta_1$ on all $\lambda_1$-paths from $z_1$ to $z_2$.
We then reduce $\lambda$ below $\lambda_1$ to a value $\lambda_2$ below
which either $z_1$ or $z_2$ is disconnected from $\zeta_1$.
This will give us a new marked point $\zeta_2$ labelled with a $\lambda$
equal to $\lambda_2$.
The procedure continues iteratively until the points $\zeta_i$ fill in a
continuous path between $z_1$ and $z_2$.

The procedure outlined above is the continuum version of a standard minimax
algorithm (see, e.g., \cite{alexander} and references therein) to construct
the minimal spanning tree on ${\mathbb Z}^2$ (using uniform $(0,1)$ bond
variables) where one looks at the minimum over all paths from $z_1$ to
$z_2$ of $\max_{b \in \text{path}}U_b$ to get a particular bond, and
then the procedure is repeated iteratively as above.

We note that the minimax value of $\lambda$ for the connection between points
in two different white (open) clusters (of the critical model) will be positive,
while the minimax value of $\lambda$ for the connection between points in the
same white (open) cluster will be negative.
The minimal spanning tree path between two points in the same continuum
white/open cluster is obtained by decreasing $\lambda$ from $0$ towards
$-\infty$, and the minimax points will be either double points of the
counterclockwise $0$-loop surrounding the cluster or points in the interior
of that counterclockwise $0$-loop where two clockwise daughter $0$-loops
touch each other or points where one such daughter loop touches the
original $0$-loop.

We remark that we have presented our continuum minimax construction
of the continuum MST in a relatively simple version that does not use
any cutoffs (like those discussed at the end of Section~\ref{connectivity}).
Even if such cutoffs turn out to be needed, the resulting construction
should still be feasible within our framework of loops and marked double
points.

\bigskip
\bigskip

\noindent {\bf Acknowledgements.}
The research of the authors was supported in part by the following sources:
for F.~C., a Marie Curie Intra-European Fellowship under contract
MEIF-CT-2003-500740 and a Veni grant of the Dutch Science Organization (NWO);
for L.~R.~G.~F., FAPESP project no.~04/07276-2 and CNPq project no.~300576/92-7;
for C.~M.~N., grant DMS-01-04278 of the U.S. NSF.
This paper has benefitted from the hospitality shown to various of
the authors at a number of venues where the research and writing
took place, including the Courant Institute of Mathematical Sciences,
the Ninth Brazilian School of Probability at Maresias Beach, Instituto
de Matem\'atica e Estat\'istica - USP, and Vrije Universiteit Amsterdam.
The authors thank Marco Isopi and Jeff Steif for useful discussions.

\end{document}